\documentclass[a4paper]{report}
\usepackage[utf8]{inputenc}
\usepackage[T1]{fontenc}
\usepackage{RJournal}
\usepackage{amsmath,amssymb,array}
\usepackage{booktabs}

\usepackage{multirow}
\usepackage{graphicx}
\usepackage{amssymb}
\usepackage{adjustbox}
\usepackage{float}
\newcommand{\RR}{\mathbb{R}}
\newcommand{\EE}{\mathbb{E}}

\newcommand{\G}{\mathrm{G}}
\newcommand{\xnew}{x_{\rm new}}
\newcommand{\argmin}{\arg\min}

\begin{document}

\sectionhead{Contributed research article}
\volume{XX}
\volnumber{YY}
\year{20ZZ}
\month{AAAA}

\begin{article}
\title{SIHR: Statistical Inference in High-Dimensional Linear and Logistic Regression Models}
\author{by Prabrisha Rakshit, Zhenyu Wang, T. Tony Cai, and Zijian Guo}

\maketitle
  
\abstract{
We introduce the R package \CRANpkg{SIHR} for statistical inference in high-dimensional generalized linear models with continuous and binary outcomes. The package provides functionalities for constructing confidence intervals and performing hypothesis tests for low-dimensional objectives in both one-sample and two-sample regression settings. We illustrate the usage of \CRANpkg{SIHR} through numerical examples and present real data applications to demonstrate the package's performance and practicality.
}
 
\section{Introduction} \label{sec:intro}
In many applications, it is common to encounter regression problems where the number of covariates $p$ exceeds the sample size $n$. Much progress has been made in point estimation and support recovery in high-dimensional generalized linear models (GLMs), as evidenced by works such as \cite{buhlmann2011statistics, negahban2009unified, huang2012estimation, lasso, scad, mcp, slasso, sqlasso, srecov}. In particular,  \citet{van2014asymptotically, javanmard2014confidence, zhang2014confidence} have proposed methods to correct the bias of penalized regression estimators and construct confidence intervals  (CIs) for individual regression coefficients of the high-dimensional linear model. Furthermore, \citet{cai2017confidence} studied the minimaxity and adaptivity of CIs for linear functionals of the regression vector in high-dimensional linear models, while \citet{cai2021statistical} proposed CIs and simultaneous hypothesis tests for individual regression coefficients in high-dimensional binary GLMs with general link functions. This debiased approach has sparked a rapidly growing research area focused on CI construction and hypothesis testing for low-dimensional objectives in high-dimensional GLMs.

In the current paper, we present the R package \CRANpkg{SIHR}, which builds on the debiasing inference method and targets a wide range of inference problems in high-dimensional GLMs for both continuous and binary outcomes. We consider the high-dimensional GLMs: for $1\leq i\leq n$,
\begin{equation}
    \mathbb{E}(y_i \mid X_{i\cdot}) = f(X_{i\cdot}^\intercal \beta),\quad \textrm{with}\;
    f(z) = \begin{cases}
        z & \quad \textrm{for linear model;}\\
        \exp{(z)} / \left[1 + \exp{(z)} \right] & \quad \textrm{for logistic model;} \\
    \end{cases}
    \label{eq: glm}
\end{equation}
where $\beta \in \mathbb{R}^p$ denotes the high-dimensional regression vector, $y_i \in \mathbb{R}$ and $X_{i\cdot} \in \mathbb{R}^p$ denote respectively the outcome and the measured covariates of the $i$-th observation. Throughout the paper, define $\Sigma = \mathbb{E}X_{i\cdot} X_{i\cdot}^\intercal$ and assume $\beta$ to be a sparse vector with its sparsity level denoted as $\|\beta\|_0$. In addition to the one-sample setting, we examine the statistical inference methods for the two-sample regression models. Particularly, we generalize the regression model in \eqref{eq: glm} and consider:
\begin{equation}
    \EE(y_i^{(k)} \mid X_{i\cdot}^{(k)}) = f(X_{i\cdot}^{(k)\intercal} \beta^{(k)}) \quad \textrm{with}\; k=1,2 \; \textrm{and}\; 1\leq i\leq n_k
    \label{eq: two sample glm}
\end{equation}
where $f(\cdot)$ is the pre-specified link function defined as \eqref{eq: glm}, $\beta^{(k)} \in \RR^p$ denotes the high-dimensional regression vector in $k$-th sample, $y_i^{(k)} \in \RR$ and $X_{i\cdot}^{(k)} \in \RR^p$ denote respectively the outcome and the measured covariates in the $k$-th sample.

The R package \CRANpkg{SIHR} consists of five main functions \code{LF()}, \code{QF()}, \code{CATE()}, \code{InnProd()}, and \code{Dist()} implementing the statistical inferences for five different quantities correspondingly, under the one-sample model \eqref{eq: glm} or two-sample model \eqref{eq: two sample glm}.
\begin{enumerate}
    \item \code{LF()}, abbreviated for linear functional, implements the inference approach for $\xnew^\intercal \beta$ proposed in \citet{cai2021optimal, cai2021statistical}, with $\xnew \in \RR^p$ denoting a loading vector.
    With $\xnew = e_j$ as a special case, \code{LF()} infers the regression coefficient $\beta_j$ \citep[e.g.]{van2014asymptotically, javanmard2014confidence, zhang2014confidence}. When $\xnew$ denotes a future observation's covariates, \code{LF()} makes inferences for the conditional mean of the outcome for the individual. See the usage of \code{LF()} in the section \nameref{subsec: LF}. 
    \item \code{QF()}, abbreviated for quadratic functional, makes inferences for $\beta_{\mathrm{G}}^{\intercal} A \beta_{\mathrm{G}}$, following the proposal in \citet{guo2019optimal, guo2021group, cai2020semisupervised}. $A\in \mathbb{R}^{|\mathrm{G}|\times |\mathrm{G}|}$ is either a pre-specified submatrix or the unknown $\Sigma_{\G,\G}$ and $\textrm{G} \in \{1,...,p\}$ denotes the index set of interest;  $\beta_{\mathrm{G}}^{\intercal} A \beta_{\mathrm{G}}$ can be viewed as a total measure of all effects of variables in the group $\mathrm{G}$. See the section \nameref{subsec: QF} for the usage. 

    \item \code{CATE()}, abbreviated for conditional average treatment effect, is to make inference for $f(\xnew^\intercal \beta^{(2)}) - f(\xnew^\intercal \beta^{(1)})$, see \cite{cai2021optimal} for detailed discussion. This difference measures the discrepancy between conditional means, closely related to the conditional average treatment effect for the new observation with covariates $\xnew$. We demonstrate its usage in the section \nameref{subsec: CATE}. 

    \item \code{InnProd()}, abbreviated for inner products, implements the statistical inference for $\beta^{(1)\intercal}_\mathrm{G} A \beta^{(2)}_\mathrm{G}$ with $A\in \R^{|G|\times |G|}$, which was proposed in \cite{guo2019optimal, ma2022statistical}. The inner products measure the similarity between the high-dimensional vectors $\beta^{(1)}$ and $\beta^{(2)}$, which is useful in capturing the genetic relatedness in the GWAS applications \citep{guo2019optimal, ma2022statistical}. 
    The usage is detailed in the section \nameref{subsec: InnProd}.
 
    \item \code{Dist()}, short-handed for distance, makes inferences for the weighted distances  $\gamma_\mathrm{G}^\intercal A \gamma_\mathrm{G}$ with $\gamma = \beta^{(2)} - \beta^{(1)}$. The distance measure is useful in comparing different high-dimensional regression vectors and constructing a generalizable model in the multisource learning problem \citet{guo2023robust}. See the section \nameref{subsec: Dist} for its usage. 
\end{enumerate}


There are a few other R packages for high-dimensional inference. The package \CRANpkg{hdi} and the package \CRANpkg{SSLASSO} implement the coordinate debiased Lasso estimators proposed in \citet{van2014asymptotically} and \citet{javanmard2014confidence} respectively. These functions provide debiased estimators of $\beta$ along with their standard error estimators, enabling confidence interval construction and hypothesis testing. However, these packages may suffer from heavy computational burden as multiple debiased optimizations are needed for inference on the linear functional.
In contrast, our R package \CRANpkg{SIHR} is computationally efficient as it performs debiasing only once. Additionally, \CRANpkg{SIHR} targets a broader range of inference targets, including the single regression coefficient as a special case.
The \CRANpkg{DoubleML} package focuses on estimating low-dimensional parameters of interest, such as causal or treatment effect parameters, in the presence of high-dimensional nuisance parameters that can be estimated using machine learning methods, while our package aims to estimate arbitrary linear and weighted quadratic combinations of the coefficient vector in high-dimensional regression.
Selective inference is implemented by the R package \CRANpkg{selectiveInference}. They focus on parameters based on the selected model, while we focus on fixed parameters independent of the selected models. 
The method proposes a one-step estimator starting from the initial LASSO estimator based on the KKT conditions for the submodel selected by LASSO. 

{
In the remainder of this paper, we provide a review of the inference methods in Section \nameref{sec: MB}, and introduce the main functions of the package in Section \nameref{sec: Package}, accompanied by illustrative examples. Finally, we demonstrate the application of our proposed methods to real data in Section \nameref{sec: real data}.
}

\section{Methodological Background}
\label{sec: MB}
We briefly review the penalized maximum likelihood estimator of $\beta$ in the high-dimensional GLM \eqref{eq: glm}, defined as:
\begin{equation}
    \widehat{\beta} = \argmin_{\beta \in \RR^p} \ell(\beta) + \lambda_0 \sum_{j=2}^p \frac{\|X_{\cdot j\|_2}}{\sqrt{n}} |\beta_j|
    \label{eq: initial estimator}
\end{equation}
with $X_{\cdot j}$ denoting the $j$-th column of $X$, the first column of $X$ set as the constant 1, and 
\begin{equation}
    \ell(\beta) = 
    \begin{cases}
        \frac{1}{n} \sum_{i=1} \left(y_i - X_{i\cdot}^{\intercal} \beta\right)^2 & \quad \textrm{for linear model}\\ 
        -\frac{1}{n} \sum_{i=1}^n y_i \log{\left[\frac{f(X_{i\cdot}^\intercal \beta)}{1 - f(X_{i\cdot}^\intercal \beta)} \right]} - \frac{1}{n} \sum_{i=1}^n \log{\left( 1 - f(X_{i\cdot}^\intercal \beta) \right)} & \quad \textrm{for GLM with binary outcome}
    \end{cases}.
\end{equation}
The tuning parameter $\lambda_0 \asymp \sqrt{\log p/n}$ is chosen by cross-validation. In the penalized regression \eqref{eq: initial estimator}, we do not penalize the intercept coefficient $\beta_1$. The penalized estimators have been shown to achieve the optimal convergence rates and satisfy desirable variable selection properties \citep{meinshausen2006high, bickel2009simultaneous, zhao2006model, wainwright2009sharp}. However, these estimators are not ready for statistical inference due to the non-negligible estimation bias induced by the penalty term \citep{van2014asymptotically, javanmard2014confidence, zhang2014confidence}. 

In section \nameref{subsec: LF-unified}, we propose a unified inference method for  $\xnew^\intercal \beta$ under linear and logistic outcome models. 
We also discuss inferences for quadratic functionals $\beta_\mathrm{G}^\intercal A \beta_{\textrm{G}}$ and $\beta_{\textrm{G}}^\intercal \Sigma_{{\textrm{G,G}}} \beta_{\textrm{G}}$ in section \nameref{subsec: QF-unified}. In the case of the two-sample high-dimensional regression model \eqref{eq: two sample glm}, we develop the inference method for conditional treatment effect $\Delta(\xnew) = f(\xnew^{\intercal} \beta^{(2)}) - f(\xnew^{\intercal} \beta^{(1)})$ in section \nameref{subsec: CATE-unified}; we consider inference for $\beta_\G^{(1)\intercal}A\beta_\G^{(2)}$ and $\beta_\G^{(1)\intercal}\Sigma_{\G,\G}\beta_\G^{(2)}$ in section \nameref{subsec: inner product} and 
$\gamma_{\G}^\intercal A \gamma_{\G}$ and $\gamma_{\G}^\intercal \Sigma_{\G,\G} \gamma_{\G}$ with $\gamma = \beta^{(2)} - \beta^{(1)}$ in section \nameref{subsec: distance}.

\subsection{Linear functional for linear model}
\label{subsec: LF-linear}
To illustrate the idea of constructing the inference method, we start with the linear functional for the linear model, which will be extended to a unified version in the section \nameref{subsec: LF-unified}. For the linear model in \eqref{eq: glm}, we define $\epsilon_i = y_i - X_{i\cdot}^\intercal \beta$ and rewrite the model as $y_i = X_{i\cdot}^\intercal \beta + \epsilon_i$ for $1\leq i\leq n$. Given the vector $\xnew \in \RR^p$, we construct the point estimator and the CI for $\xnew^\intercal \beta$.

A natural idea for the point estimator is to use the plug-in estimator $\xnew^\intercal \widehat{\beta}$ with the penalized estimator $\widehat{\beta}$ defined in \eqref{eq: initial estimator}. However,  the bias $\xnew^\intercal (\widehat{\beta} -\beta)$ is not negligible. The work \citet{cai2021optimal} proposed the bias-corrected estimator as,
\begin{equation}
    \widehat{\xnew^\intercal \beta} = \xnew^\intercal \widehat{\beta} +
    \widehat{u}^\intercal \frac{1}{n} \sum_{i=1}^n X_{i\cdot} \left(y_i - X_{i\cdot}^\intercal \widehat{\beta}\right)
    \label{eq: LF point estimator - linear}
\end{equation}
where the second term on the right hand side in \eqref{eq: LF point estimator - linear} is the estimate of negative bias $- \xnew^\intercal (\widehat{\beta}-\beta)$, 
and the projection direction $\widehat{u}$ is defined as
\begin{align}
    \widehat{u} = \argmin_{u \in \mathbb{R}^p} u^\intercal \widehat{\Sigma} u \quad \textrm{ subject to: }
    &\; \|\widehat{\Sigma} u - \xnew\|_\infty \leq \|\xnew\|_2 \lambda 
    \label{eq: proj_direc - linear}\\
    &\; \left\lvert \xnew^\intercal \widehat{\Sigma} u - \|\xnew\|^2_2 \right\rvert \leq \|\xnew\|^2_2 \lambda \label{eq: proj_direc - linear - constraint2}
\end{align}
where $\widehat{\Sigma} = \frac{1}{n}\sum_{i=1}^n X_{i\cdot} X_{i\cdot}^\intercal$ and $\lambda \asymp \sqrt{\log p/n}$. The bias-corrected estimator $\widehat{\xnew^\intercal \beta}$ satisfies the following error decomposition,
\begin{equation*}
    \widehat{\xnew^\intercal \beta} - \xnew^\intercal \beta = \underbrace{\widehat{u}^\intercal \frac{1}{n} \sum_{i=1}^n X_{i\cdot}^\intercal \epsilon_i}_\textrm{asymp. normal} + \underbrace{\left(\widehat{\Sigma} \widehat{u} - \xnew \right)^\intercal (\beta - \widehat{\beta})}_\textrm{remaining bias}
\end{equation*}
The first constraint in \eqref{eq: proj_direc - linear} controls the remaining bias term in the above equation while the second constraint in  \eqref{eq: proj_direc - linear - constraint2} is crucial to ensuring the asymptotic normality of $\widehat{\xnew^\intercal \beta} - \xnew^\intercal \beta$ for any vector $\xnew$ such that the variance of the ``asymp. normal'' term always dominates the ``remaining bias'' term. Based on the asymptotic normality, we construct the CI for $\xnew^\intercal \beta$ as 
\begin{equation*}
    \mathrm{CI}=\left(\widehat{\xnew^{\intercal}\beta}-z_{\alpha / 2} \sqrt{\widehat{\mathrm{V}}}, \quad \widehat{\xnew^{\intercal}\beta}+z_{\alpha / 2} \sqrt{\widehat{\mathrm{V}}}\right) \quad \textrm{with}\; \widehat{\mathrm{V}} = \frac{\widehat{\sigma}^2}{n} \widehat{u}^\intercal \widehat{\Sigma} \widehat{u}
\end{equation*}
where $\widehat{\sigma}^2 = \frac{1}{n}\sum_{i=1}^n (y_i - X_{i\cdot}^\intercal \widehat{\beta})^2$ and $z_{\alpha/2}$ denotes the upper $\alpha/2$ quantile for the standard normal distribution.


\subsection{Linear functional for GLM}
\label{subsec: LF-unified}
In this subsection, we generalize the inference method specifically for the linear model in \nameref{subsec: LF-linear} to GLM in \eqref{eq: glm}. Given the initial estimator $\widehat{\beta}$, the key step is to estimate the bias $\xnew^\intercal (\widehat{\beta} - \beta)$. 
We can propose a unified version of the bias-corrected estimator for $\xnew^\intercal \beta$ as
\begin{equation}
    \widehat{\xnew^\intercal \beta} = \xnew^\intercal \widehat{\beta} +
    \widehat{u}^\intercal \frac{1}{n} \sum_{i=1}^n \omega(X_{i\cdot}^\intercal \widehat{\beta}) \left(y_i - f(X_{i\cdot}^\intercal \widehat{\beta})\right) X_{i\cdot}
    \label{eq: LF point estimator - unified}
\end{equation}
with the second term on the right hand side of \eqref{eq: LF point estimator - unified} being the estimate of $-\xnew^\intercal (\widehat{\beta} - \beta)$. In consideration of different link functions $f(\cdot)$ in \eqref{eq: glm}, we shall specify in the following how to construct the projection direction $\hat{u}$ and the weight function $\omega: \mathbb{R} \mapsto \mathbb{R}$ in \eqref{eq: LF point estimator - unified}.
\begin{table}[ht]
    \centering
    \resizebox{0.75\linewidth}{!}{
    \begin{tabular}{|l|l|c|c|c|l|}
        \hline
        Model & Outcome Type & $f(z)$ & $f^\prime(z)$ & $\omega(z)$ & Weighting\\
        \hline
        linear & Continuous & z & 1 & 1 & \\
        logistic & Binary & $\frac{e^z}{1+e^z}$ & $\frac{e^z}{(1+e^z)^2}$ & $\frac{(1+e^z)^2}{e^z}$ & Linearization\\
        logistic\_alter & Binary & $\frac{e^z}{1+e^z}$ & $\frac{e^z}{(1+e^z)^2}$ & 1 & Link-specific\\
        \hline
    \end{tabular}
    }
    \caption{Definitions of the functions $\omega$ and $f$ for different GLMs.}
    \label{tab: components}
\end{table}
In Table \ref{tab: components}, we consider different GLM models and present the corresponding functions $f(\cdot)$ and $\omega(\cdot)$, together with the derivative $f^\prime (\cdot)$.  Note that there are two ways of specifying the weights $w(z)$ for logistic regression. The linearization weighting is proposed in \citet{guo2021group} specifically for logistic regression; while \citet{cai2021statistical} constructed the link-specific weighting method for general link function $f(\cdot)$.
\noindent
The projection direction $\widehat{u} \in \RR^p$ in \eqref{eq: LF point estimator - unified} is constructed as follows:
\begin{equation}
    \begin{aligned}
        &\widehat{u} = \argmin_{u \in \mathbb{R}^p} u^\intercal \left[\frac{1}{n}\sum_{i=1}^n \omega(X_{i\cdot}^\intercal \widehat{\beta}) f^\prime (X_{i\cdot}^\intercal \widehat{\beta}) X_{i\cdot} X_{i\cdot}^\intercal \right] u \quad \textrm{ subject to: } \\
    &\quad \quad \left\|\frac{1}{n} \sum_{i=1}^n \omega(X_{i\cdot}^\intercal \widehat{\beta}) f^\prime(X_{i\cdot}^\intercal \widehat{\beta}) X_{i\cdot} X_{i\cdot}^\intercal u - \xnew \right\|_\infty \leq \|\xnew\|_2 \lambda \\
    &\quad \quad \left|\xnew^\intercal \frac{1}{n} \sum_{i=1}^n \omega(X_{i\cdot}^\intercal \widehat{\beta}) f^\prime(X_{i\cdot}^\intercal \widehat{\beta}) X_{i\cdot} X_{i\cdot}^\intercal u - \|\xnew\|^2_2 \right| \leq \|\xnew\|_2^2 \lambda .
    \end{aligned}
    \label{eq: projection}
\end{equation}
It has been established that $\widehat{\xnew^\intercal \beta}$ in \eqref{eq: LF point estimator - unified} is asymptotically unbiased and normal for the linear model \citep{cai2021optimal}, the logistic model \citep{guo2021inference,cai2021statistical}, and the probit model \citep{cai2021statistical}. The variance of $\widehat{\xnew^\intercal \beta}$ can be estimated by $\widehat{\mathrm{V}}$, defined as
\begin{align}
    \widehat{\mathrm{V}} &= \widehat{u}^\intercal \left[\frac{1}{n^2} \sum_{i=1}^n \left(\omega(X_{i\cdot}^\intercal \widehat{\beta}) \right)^2 \widehat{\sigma}_i^2 X_{i\cdot} X_{i\cdot}^\intercal \right] \widehat{u}
    \quad \textrm{with}: \; \label{eq: LF Variance - unified}\\
    &\quad \widehat{\sigma}_i^2 = 
    \begin{cases}
    \frac{1}{n} \sum_{j=1}^n \left(y_j - X_{j\cdot}^\intercal \widehat{\beta}\right)^2, &\textrm{for linear model} \vspace{0.2cm}\\
    f(X_{i\cdot}^\intercal \widehat{\beta}) (1 - f(X_{i\cdot}^\intercal \widehat{\beta})), &\textrm{for GLM with binary outcome}
    \end{cases}.
    \label{eq: noise level - unified}
\end{align}
Based on the asymptotic normality, the CI for $\xnew^\intercal \beta$ is:
\begin{equation*}
\mathrm{CI}=\left(\widehat{\xnew^{\intercal}\beta}-z_{\alpha / 2} \sqrt{\widehat{\mathrm{V}}}, \quad \widehat{\xnew^{\intercal}\beta}+z_{\alpha / 2} \sqrt{\widehat{\mathrm{V}}}\right).
\label{eq: LF CI - unified}
\end{equation*}
Subsequently, for the binary outcome case, we estimate the case probability $\mathbb{P}(y_i = 1 \mid X_{i\cdot} = \xnew)$ by $f(\widehat{\xnew^\intercal \beta})$ and construct the CI for $f(\xnew^\intercal \beta)$ as:
\begin{equation*}
    \mathrm{CI} = \left(f\left(\widehat{\xnew^\intercal \beta} - z_{\alpha /2}\sqrt{\widehat{\mathrm{V}}} \right), 
    f\left(\widehat{\xnew^\intercal \beta} + z_{\alpha /2}\sqrt{\widehat{\mathrm{V}}} \right)\right).
\end{equation*}

\subsection{Quadratic functional for GLM}
\label{subsec: QF-unified}
We now move our focus to inference for the quadratic functional $\mathrm{Q}_A =\beta_{\mathrm{G}}^\intercal A \beta_{\mathrm{G}}$, where $G \subset \{1,..., p\}$ and $A \in \RR^{|G|\times |G|}$ denotes a pre-specified matrix of interest. Without loss of generality, we set $G=\{1,2,\cdots,|G|\}$. In the following, we propose a unified version of the point estimator and CI under the GLM \eqref{eq: glm}. With the initial estimator $\widehat{\beta}$ defined in \eqref{eq: initial estimator}, the plug-in estimator $\widehat{\beta}_{\mathrm{G}}^\intercal A \widehat{\beta}_{\mathrm{G}}$ suffers from the following error,
\begin{equation*}
    \widehat{\beta}_{\textrm{G}}^\intercal A \widehat{\beta}_{\textrm{G}} - \beta_{\textrm{G}}^\intercal A \beta_{\textrm{G}} = 2 \widehat{\beta}_{\textrm{G}}^\intercal A ( \widehat{\beta}_{\textrm{G}} - \beta_{\textrm{G}} ) - ( \widehat{\beta}_{\textrm{G}} - \beta_{\textrm{G}} )^\intercal A ( \widehat{\beta}_{\textrm{G}} - \beta_{\textrm{G}} ).
\end{equation*}
The last term in the above decomposition $( \widehat{\beta}_{\textrm{G}} - \beta_{\textrm{G}} )^\intercal A ( \widehat{\beta}_{\textrm{G}} - \beta_{\textrm{G}} )$ is the higher-order approximation error under regular conditions; thus the bias mainly comes from the term $2 \widehat{\beta}_{\textrm{G}}^\intercal A ( \widehat{\beta}_{\textrm{G}} - \beta_{\textrm{G}} )$, which can be expressed as $2 \,\xnew^\intercal (\widehat{\beta} - \beta)$ with $\xnew = ( \widehat{\beta}_{\textrm{G}}^\intercal A, \; \mathbf{0})^\intercal$. Hence the term can be estimated directly by applying the linear functional approach in section \nameref{subsec: LF-unified}. Utilizing this idea, \citet{guo2021group, guo2019optimal} proposed the following estimator of $\mathrm{Q}_A$,
\begin{equation*}
    \widehat{\mathrm{Q}}_A = \widehat{\beta}_{\textrm{G}}^\intercal A \widehat{\beta}_{\textrm{G}} + 2\, \widehat{u}_A^\intercal \left[ \frac{1}{n} \sum_{i=1}^n \omega(X_{i\cdot}^\intercal \widehat{\beta}) \left(y_i - f(X_{i\cdot}^\intercal \widehat{\beta})\right) X_{i\cdot} \right] 
\end{equation*}
with the second term being the estimate of $-2\widehat{\beta}_{\textrm{G}}^\intercal A ( \widehat{\beta}_{\textrm{G}} - \beta_{\textrm{G}} )$, where $\widehat{u}_A$ is the projection direction defined in \eqref{eq: projection} with $\xnew = ( \widehat{\beta}_{\textrm{G}}^\intercal A, \; \mathbf{0}^{\intercal})^\intercal$.
Since ${\mathrm{Q}}_A$ is non-negative if $A$ is positive semi-definite, we truncate $\widehat{\mathrm{Q}}_A$ at $0$ and define
$
\widehat{\mathrm{Q}}_A = \max\left(\widehat{\mathrm{Q}}_A, \; 0\right)
$
We further estimate the variance of the $\widehat{\mathrm{Q}}_A$ by
\begin{equation}
    \widehat{\mathrm{V}}_A(\tau) = 4 \widehat{u}^\intercal \left[\frac{1}{n^2} \sum_{i=1}^n \omega^2(X_{i\cdot}^\intercal \widehat{\beta}) \widehat{\sigma}_i^2 X_{i\cdot} X_{i\cdot}^\intercal \right] \widehat{u} + \frac{\tau}{n}
    \label{eq: QF CI1 - unifed}
\end{equation}
where the term $\tau/n$ with $\tau>0$ (default value $\tau=1$) is introduced as an upper bound for the term $( \widehat{\beta}_{\textrm{G}} - \beta_{\textrm{G}} )^\intercal A ( \widehat{\beta}_{\textrm{G}} - \beta_{\textrm{G}} )$, and $\widehat{\sigma}^2_i$ is defined in \eqref{eq: noise level - unified}. 
Then given a fixed value of $\tau$, we construct the CI as 
${\rm CI}(\tau) = \left(\max\left(\widehat{\rm Q}_A - z_{\alpha/2}\sqrt{\widehat{\rm V}_A(\tau)},\; 0\right),\; \widehat{\rm Q}_A + z_{\alpha/2}\sqrt{\widehat{\rm V}_A(\tau)}\right).$ 

Now we turn to the estimation of $\mathrm{Q}_{\Sigma} = \beta_{\textrm{G}}^\intercal \Sigma_{{\textrm{G,G}}} \beta_{\textrm{G}}$ where the matrix $\Sigma_{{\textrm{G,G}}}$ is unknown and  estimated by $\widehat{\Sigma}_{\G,\G} = \frac{1}{n}\sum_{i=1}^{n}X_{i\G}X_{i\G}^{\intercal}$. 
Decompose the error of the plug-in estimator $\widehat{\beta}_{\textrm{G}}^\intercal \widehat{\Sigma}_{\textrm{G,G}} \widehat{\beta}$:
\begin{equation*}
    \widehat{\beta}_{\textrm{G}}^\intercal \widehat{\Sigma}_{\textrm{G,G}} \widehat{\beta} - \beta_{\textrm{G}} \Sigma_{{\textrm{G,G}}} \beta_{\textrm{G}} = 2\, \widehat{\beta}_{\textrm{G}}^\intercal \widehat{\Sigma}_{\textrm{G,G}} (\widehat{\beta}_{\textrm{G}} - \beta_{\textrm{G}}) + \beta_{\textrm{G}}^\intercal (\widehat{\Sigma}_{\textrm{G,G}} - \Sigma_{{\textrm{G,G}}})\beta_{\textrm{G}} - (\widehat{\beta}_{\textrm{G}} - \beta_{\textrm{G}})^\intercal \widehat{\Sigma}_{\textrm{G,G}} (\widehat{\beta}_{\textrm{G}} - \beta_{\textrm{G}}).
\end{equation*}
The first term $\widehat{\beta}_{\textrm{G}}^\intercal \widehat{\Sigma}_{\textrm{G,G}} (\widehat{\beta}_{\textrm{G}} - \beta_{\textrm{G}})$ is estimated by applying linear functional approach in \nameref{subsec: LF-unified} with $\xnew = ( \widehat{\beta}_{\textrm{G}}^\intercal \widehat{\Sigma}_{\G, \G}, \; \mathbf{0})^\intercal$; 
the second term $\beta_{\textrm{G}}^\intercal (\widehat{\Sigma}_{\textrm{G,G}} - \Sigma_{{\textrm{G,G}}})\beta_{\textrm{G}}$ can be controlled asymptotically by central limit theorem;
and the last term $(\widehat{\beta}_{\textrm{G}} - \beta_{\textrm{G}})^\intercal \widehat{\Sigma}_{\textrm{G,G}} (\widehat{\beta}_{\textrm{G}} - \beta_{\textrm{G}})$ is negligible due to high-order bias. \citet{guo2021group} proposed the following estimator of $\mathrm{Q}_\Sigma$
\begin{equation*}
    \widehat{\mathrm{Q}}_{\Sigma} = \widehat{\beta}_{\textrm{G}}^\intercal \widehat{\Sigma}_{\textrm{G,G}} \widehat{\beta}_{\textrm{G}} + 2\, \widehat{u}_\Sigma^\intercal \left[ \frac{1}{n} \sum_{i=1}^n \omega(X_{i\cdot}^\intercal \widehat{\beta}) \left(y_i - f(X_{i\cdot}^\intercal \widehat{\beta})\right) X_{i\cdot} \right]  
\end{equation*}
where $\widehat{u}_\Sigma$ is the projection direction constructed in \eqref{eq: projection} with $\xnew = (\widehat{\beta}^\intercal_{\textrm{G}} \widehat{\Sigma}_{\textrm{G,G}}, \; \mathbf{0})^\intercal$. We introduce the estimator $\widehat{\rm Q}_\Sigma = \max(\widehat{\rm Q}_\Sigma,\; 0)$ and estimate its variance as 
\begin{equation}
    \widehat{\mathrm{V}}_\Sigma(\tau) = 4 \widehat{u}^\intercal \left[\frac{1}{n^2} \sum_{i=1}^n \omega^2(X_{i\cdot}^\intercal \widehat{\beta}) \widehat{\sigma}_i^2 X_{i\cdot} X_{i\cdot}^\intercal \right] \widehat{u} + 
    \frac{1}{n^2} \sum_{i=1}^n \left(\widehat{\beta}^\intercal_{\textrm{G}} X_{i,G} X_{i,G}^\intercal \widehat{\beta}_{\textrm{G}} - \widehat{\beta}_{\textrm{G}}^\intercal \widehat{\Sigma}_{\textrm{G,G}}\widehat{\beta}_{\textrm{G}} \right)^2
    +\frac{\tau}{n}
    \label{eq: QF CI2 - unified}
\end{equation}
where $\tau > 0$, the term $\tau/n$ is introduced as an upper bound for the term $( \widehat{\beta}_{\textrm{G}} - \beta_{\textrm{G}} )^\intercal \widehat{\Sigma}_{\G,\G} ( \widehat{\beta}_{\textrm{G}} - \beta_{\textrm{G}} )$, and $\widehat{\sigma}^2_i$ is defined in \eqref{eq: noise level - unified}.
{Then, thanks to the asymptotic normality, for a fixed value of $\tau$, we can construct the CI as 
$${\rm CI}(\tau) = \left(\max\left(\widehat{\rm Q}_{\Sigma} - z_{\alpha/2}\sqrt{\widehat{\rm V}_{\Sigma}(\tau)}, \; 0\right),\; \widehat{\rm Q}_{\Sigma} + z_{\alpha/2}\sqrt{\widehat{\rm V}_{\Sigma}(\tau)}\right)$$ }

\subsection{Conditional average treatment effects}
\label{subsec: CATE-unified}

The inference methods proposed for one sample can be generalized to make inferences for conditional average treatment effects, which can be expressed as the difference between two linear functionals. Let $A_i \in \{1,2\}$ denote the treatment assignment for $i$-th observation. 
Consider the two-sample GLMs as
\begin{equation*}
    \EE(y_i | X_{i\cdot}, A_i = 1) = f(X_{i\cdot}^\intercal \beta^{(1)})
    \quad \textrm{and} \quad
    \EE(y_i | X_{i\cdot}, A_i = 2) = f(X_{i\cdot}^\intercal \beta^{(2)})
\end{equation*}
where $f$ is the link function listed in table \ref{tab: components} 
Then, for a future individual $X_{i\cdot} = \xnew$, we define $\Delta(\xnew) = \EE(y_i | X_{i\cdot}, A_i = 2) - \EE(y_i | X_{i\cdot}, A_i = 1)$, that measures the difference of the conditional mean of assignment of treatment for the individual with covariates $\xnew$.

Following \eqref{eq: LF point estimator - unified}, we construct the bias-corrected point estimators of $\widehat{\xnew^\intercal \beta^{(1)}}$ and $\widehat{\xnew^\intercal \beta^{(2)}}$, together with their corresponding variance $\widehat{\mathrm{V}}_{(1)}$ and $\widehat{\mathrm{V}}_{(2)}$ as \eqref{eq: LF Variance - unified}. The paper \citet{cai2021optimal} proposed to estimate $\Delta(\xnew)$ by $\widehat{\Delta}(\xnew)$ as:
\begin{equation*}
    \widehat{\Delta}(\xnew) = f(\widehat{\xnew^\intercal \beta^{(2)}}) - f(\widehat{\xnew^\intercal \beta^{(1)}})
\end{equation*}
Its variance can be estimated with delta method by:
\begin{equation*}
    \widehat{\mathrm{V}}_\Delta = \left(f^\prime(\widehat{\xnew^\intercal \beta^{(1)}})\right)^2  \widehat{\mathrm{V}}_{(1)} + \left(f^\prime(\widehat{\xnew^\intercal \beta^{(2)}})\right)^2  \widehat{\mathrm{V}}_{(2)}
\end{equation*}
{ Then we construct the CI as
 ${\rm CI} = \left(\widehat{\Delta}(\xnew) - z_{\alpha/2}\sqrt{\widehat{\rm V}_{\Delta}}, \widehat{\Delta}(\xnew) + z_{\alpha/2}\sqrt{\widehat{\rm V}_{\Delta}}\right).$
 }



\subsection{Inner product of regression vectors}
\label{subsec: inner product}
The paper \citet{guo2019optimal, ma2022statistical} have carefully investigated the CI construction for $\beta^{(1)\intercal}_\G A \beta^{(2)}_\G$, provided with a pre-specified submatrix $A \in \RR^{|\mathrm{G}|\times |\mathrm{G}|}$ and the set of indices $\mathrm{G} \in \{1,..., p\}$. Let $\widehat{\beta}^{(1)}$ and $\widehat{\beta}^{(2)}$ respectively be the initial estimators for their corresponding sample in \eqref{eq: two sample glm}, the plug-in but biased estimator is $\widehat{\beta}^{(1)\intercal}_\mathrm{G} A \widehat{\beta}^{(2)}_\mathrm{G}$. Its bias can be decomposed as:
\begin{multline*}
    \widehat{\beta}^{(1)\intercal}_\mathrm{G} A \widehat{\beta}^{(2)}_\mathrm{G} - \beta_\mathrm{G}^{(1)\intercal} A \beta_\mathrm{G}^{(2)} = \widehat{\beta}_\G^{(2)\intercal} A \left(\widehat{\beta}_\G^{(1)} - \beta_\G^{(1)}\right) + \widehat{\beta}_\G^{(1)\intercal} A \left(\widehat{\beta}_\G^{(2)} - \beta_\G^{(2)}\right)  \\
    - \left(\widehat{\beta}_\G^{(1)} - \beta_\G^{(1)}\right)^\intercal A \left(\widehat{\beta}_\G^{(2)} - \beta_\G^{(2)}\right).
\end{multline*}
The key step is to estimate the error components $ \widehat{\beta}_\G^{(2)\intercal} A \left(\widehat{\beta}_\G^{(1)} - \beta_\G^{(1)}\right) $ and $\widehat{\beta}_\G^{(1)\intercal} A \left(\widehat{\beta}_\G^{(2)} -  \beta_\G^{(2)}\right)$. Then the following procedures can be interpreted as applying Linear Functional twice on two independent samples. To be specific, 
we propose the following bias-corrected estimator for $\beta_\G^{(1)\intercal} A \beta_\G^{(2)}$
\begin{equation}
    \begin{aligned}
        \widehat{\beta_\G^{(1)\intercal}A\beta_\G^{(2)}} = \widehat{\beta}_\G^{(1)\intercal}A\widehat{\beta}_\G^{(2)} + & \widehat{u}_1^{\intercal}\frac{1}{n_1}\sum_{i=1}^{n_1}\omega(X_{i\cdot}^{(1)\intercal}\widehat{\beta}^{(1)})\left(y_i^{(1)} - f(X_{i\cdot}^{(1)\intercal}\widehat{\beta}^{(1)})\right)X_{i\cdot}^{(1)} \\
        & + \widehat{u}_2^{\intercal}\frac{1}{n_2}\sum_{i=1}^{n_2}\omega(X_{i\cdot}^{(2)\intercal}\widehat{\beta}^{(2)})\left(y_i^{(2)} - f(X_{i\cdot}^{(2)\intercal}\widehat{\beta}^{(2)})\right)X_{i\cdot}^{(2)}
    \end{aligned}
    \label{eq: point-inner}
\end{equation}
with the second term and the third term in right-hand-side of \eqref{eq: point-inner} estimating $- \widehat{\beta}_\G^{(2)\intercal} A \left(\widehat{\beta}_\G^{(1)} - \beta_\G^{(1)}\right) $ and $- \widehat{\beta}_\G^{(1)\intercal} A \left(\widehat{\beta}_\G^{(2)} - \beta_\G^{(2)}\right)$ respectively, 
where $\widehat{u}_1$ is the projection direction defined in \eqref{eq: projection} with $\xnew = ( \widehat{\beta}_{\textrm{G}}^{(2)\intercal} A, \; \mathbf{0})^\intercal$ and $\widehat{u}_2$ is the projection direction defined in \eqref{eq: projection} with $\xnew = ( \widehat{\beta}_{\textrm{G}}^{(1)\intercal} A, \; \mathbf{0})^\intercal$. The corresponding variance of $\widehat{\beta_\G^{(1)\intercal}A\beta_\G^{(2)}}$, when $A$ is a known positive definite matrix, is estimated as
\begin{equation*}
    \widehat{\rm V}_{A}(\tau) = \widehat{\rm V}^{(1)} + \widehat{\rm V}^{(2)} + \frac{\tau}{\min(n_1,n_2)}
\end{equation*}
where $\widehat{V}^{(k)}$ is computed as \eqref{eq: LF Variance - unified} for the $k-$th regression model $(k = 1,2)$ in \eqref{eq: two sample glm} and $\tau > 0$, the term $\tau/\min(n_1,n_2)$ is introduced as an upper bound for the term $( \widehat{\beta}_{\textrm{G}}^{(1)} - \beta_{\textrm{G}}^{(1)} )^\intercal A ( \widehat{\beta}_{\textrm{G}}^{(2)} - \beta_{\textrm{G}}^{(2)} )$.

When $A$ is not specified, we treat $A = \Sigma_{\G, \G}$, which is unknown. As a natural generalization, the quantity $\beta_\mathrm{G}^{(1)\intercal} \Sigma_{\G, \G} \beta_\mathrm{G}^{(2)}$ is well defined if the two regression models in \eqref{eq: two sample glm} share the design covariance matrix $\Sigma = \EE X_{i\cdot}^{(1)} X_{i\cdot}^{(1)\intercal} = \EE X_{i\cdot}^{(2)} X_{i\cdot}^{(2)\intercal} $. We follow the above procedures replacing $A$ by $\widehat{\Sigma}_{\G,\G} =\frac{1}{n_1 + n_2} \sum_{i=1}^{n_1+n_2} X_{i,\G} X_{i, \G}^\intercal$ where $X$ is the row-combined matrix of $X^{(1)}$ and $X^{(2)}$. The variance of $\widehat{\beta_\mathrm{G}^{(1)\intercal} \Sigma_{\G, \G} \beta_\mathrm{G}^{(2)}}$ is now estimated as 
\begin{equation*}
    \widehat{\rm V}_{\Sigma}(\tau) = \widehat{\rm V}^{(1)} + \widehat{\rm V}^{(2)} + \frac{1}{(n_1 + n_2)^2}\sum_{i=1}^{n_1+n_2} \left(\widehat{\beta}_\G^{(1)\intercal}X_{i,\G}X_{i,\G}^{\intercal}\widehat{\beta}_\G^{(2)} - \widehat{\beta}_\G^{(1)\intercal}\widehat{\Sigma}_{\G,\G}\widehat{\beta}_\G^{(2)}\right)^2 + \frac{\tau}{\min(n_1,n_2)}
\end{equation*}

Depending on whether the submatrix $A$ is specified or not, the CI is
\[
{\rm CI}(\tau) = 
\begin{cases}
\left(\widehat{\beta_\G^{(1)\intercal}A\beta_\G^{(2)}} - z_{\alpha/2}\widehat{\rm V}_A(\tau), \;\; \widehat{\beta_\G^{(1)\intercal}A\beta_\G^{(2)}} + z_{\alpha/2}\widehat{\rm V}_A(\tau)\right) & \textrm{if $A$ is specified}\\
\left(\widehat{\beta_\G^{(1)\intercal}\Sigma_{\G,\G}\beta_\G^{(2)}} - z_{\alpha/2}\widehat{\rm V}_\Sigma(\tau), \;\;\widehat{\beta_\G^{(1)\intercal}\Sigma_{\G,\G}\beta_\G^{(2)}} + z_{\alpha/2}\widehat{\rm V}_\Sigma(\tau)\right) & \textrm{otherwise}
\end{cases}
\]


\subsection{Distance of regression vectors}
\label{subsec: distance}
We denote $\gamma = \beta^{(2)} - \beta^{(1)}$ and its initial estimator $\widehat{\gamma} = \widehat{\beta}^{(2)} - \widehat{\beta}^{(1)}$. The quantity of interest is the distance between two regression vectors $\gamma^\intercal_\G A \gamma_\G$, given a pre-specified submatrix $A \in \RR^{|\mathrm{G}|\times |\mathrm{G}|}$ and the set of indices $\mathrm{G} \in \{1,..., p\}$. The bias of the plug-in estimator $\widehat{\gamma}^\intercal_\G A \widehat{\gamma}_\G$ is:
\begin{equation*}
    \widehat{\gamma}^\intercal_\G A \widehat{\gamma}_\G - \gamma^\intercal_\G A \gamma_\G = 2\; \widehat{\gamma}^\intercal_\G A \left(\widehat{\beta}^{(2)}_\G - \beta^{(2)}_\G\right) - 2\; \widehat{\gamma}^\intercal_\G A \left(\widehat{\beta}^{(1)}_\G - \beta^{(1)}_\G\right) - \left(\widehat{\gamma}_\G - \gamma_\G\right)^\intercal A \left(\widehat{\gamma}_\G - \gamma_\G\right)
\end{equation*}
The key step is to estimate the error components $\widehat{\gamma}_\G^\intercal A \left(\widehat{\beta}_\G^{(1)} -  \beta_\G^{(1)}\right)$ and $ \widehat{\gamma}_\G^\intercal A \left(\widehat{\beta}_\G^{(2)} - \beta_\G^{(2)}\right) $ in the above decomposition. We apply linear functional techniques twice here, and propose the bias-corrected estimator:
\begin{equation}
    \begin{aligned}
        \widehat{\gamma_\G^\intercal A \gamma_\G} = 
        \widehat{\gamma}_\G^\intercal A \widehat{\gamma}_\G & - 2\,\widehat{u}_1^{\intercal}\frac{1}{n_1}\sum_{i=1}^{n_1}\omega(X_{i\cdot}^{(1)\intercal}\widehat{\beta}^{(1)})\left(y_i^{(1)} - f(X_{i\cdot}^{(1)\intercal}\widehat{\beta}^{(1)})\right)X_{i\cdot}^{(1)} \\
        &+ 2\,\widehat{u}_2^{\intercal}\frac{1}{n_2}\sum_{i=1}^{n_2}\omega(X_{i\cdot}^{(2)\intercal}\widehat{\beta}^{(2)})\left(y_i^{(2)} - f(X_{i\cdot}^{(2)\intercal}\widehat{\beta}^{(2)})\right)X_{i\cdot}^{(2)}
    \end{aligned}
    \label{eq: point-distance}
\end{equation}
Then by non-negative distance, we define 
$
\widehat{\gamma_\G^\intercal A \gamma_\G} = \max\left\{\widehat{\gamma_\G^\intercal A \gamma_\G}, \; 0\right\}. 
$
The second term on right-hand-side of \eqref{eq: point-distance} is to estimate $-2\, \xnew^\intercal(\widehat{\beta}_\G^{(1)} - \beta_\G^{(1)})$
with $\xnew=\left(\widehat{\gamma}_\G^\intercal A,  \mathbf{0} \right)^\intercal$; and the third term on right-hand-side of \eqref{eq: point-distance} is to estimate $-2\, \xnew^\intercal(\widehat{\beta}_\G^{(2)} - \beta_\G^{(2)})$ with $\xnew=\left(\widehat{\gamma}_\G^\intercal A,  \mathbf{0} \right)^\intercal$ as well. {The corresponding asymptotic variance for the bias-corrected estimator is
\[
\widehat{\rm V}_{A}(\tau) = 4\,\widehat{\rm V}^{(1)} + 4\,\widehat{\rm V}^{(2)} + \frac{\tau}{\min(n_1,n_2)}
\]
where $\widehat{\rm V}^{(k)}$ is computed as \eqref{eq: LF Variance - unified} for the $k$-th regression model $(k = 1,2)$ and $\tau > 0$, the term $\tau/\min(n_1,n_2)$ is introduced as an upper bound for the term $( \widehat{\gamma}_{\textrm{G}} - \gamma_{\textrm{G}} )^\intercal A ( \widehat{\gamma}_{\textrm{G}} - \gamma_{\textrm{G}})$. With asymptotic normality, we construct the CI
$${\rm CI}(\tau) = \left(\max\left(\widehat{\gamma_\G^\intercal A \gamma_\G} - z_{\alpha/2}\sqrt{\widehat{\rm V}_{A}(\tau)}, \; 0\right),\; \widehat{\gamma_\G^\intercal A \gamma_\G} + z_{\alpha/2}\sqrt{\widehat{\rm V}_{A}(\tau)}\right).$$

When the submatrix $A$ is not specified, we treat $A=\Sigma_{\G,\G}$, which is unknown. The point estimator $\widehat{\gamma_\G^\top \Sigma_{\G,\G} \gamma_\G}$ can be computed similarly as outlined in \eqref{eq: point-distance}. In this case, the submatrix $A$ is substituted with $\widehat{\Sigma}_{\G,\G}$ and the resulting value is truncated at $0$, where $\widehat{\Sigma}_{\G,\G} =\frac{1}{n_1 + n_2} \sum_{i=1}^{n_1+n_2} X_{i,\G} X_{i, \G}^\intercal$ with $X$ as the row-combined matrix of $X^{(1)}$ and $X^{(2)}$. Its corresponding asymptotic variance is
\[
\widehat{\rm V}_{\Sigma} = 4\, \widehat{\rm V}^{(1)} + 4\,\widehat{\rm V}^{(2)} + \frac{1}{(n_1 + n_2)^2}\sum_{i=1}^{n_1+n_2} \left(\widehat{\gamma}_\G^{\intercal}X_{i,\G}X_{i,\G}^{\intercal}\widehat{\gamma}_\G - \widehat{\gamma}_\G^{\intercal}\widehat{\Sigma}_{\G,\G}\widehat{\gamma}_\G\right)^2 + \frac{\tau}{\min(n_1,n_2)}
\]
Next we present its CI
$${\rm CI}(\tau) = \left(\max\left(\widehat{\gamma_\G^\intercal \Sigma \gamma_\G} - z_{\alpha/2}\sqrt{\widehat{\rm V}_{\Sigma}(\tau)}, \; 0\right), \;\widehat{\gamma_\G^\intercal \Sigma \gamma_\G} + z_{\alpha/2}\sqrt{\widehat{\rm V}_{\Sigma}(\tau)}\right)$$
}

\section{Usage of the package} \label{sec: Package}

The \CRANpkg{SIHR} package contains a set of functions for inference methods of various low-dimensional objectives, such as linear and quadratic functions. See the table \ref{tab: overview_package} for each function and its corresponding objective.
\begin{table}[htp]
    \centering
    \small
    \resizebox{0.9\textwidth}{!}{
    \begin{tabular}{m{1.5cm} m{5cm} m{8cm}}
        \toprule
        Function & Objective & Description \\
        \midrule
        LF() & $\xnew^\intercal \beta$ & Generate an LF object. \\
        QF() & $\beta_{G}^{\intercal}A\beta_{G}$  & Generate a QF object.\\
        CATE() & $f(\xnew^\intercal \beta^{(2)}) - f(\xnew^\intercal \beta^{(1)})$ & Generate a CATE object. \\
        InnProd() & $\beta^{(1)\intercal}_\G A \beta^{(2)}_\G$  & Generate an InnProd object. \\
        Dist() & $\gamma^\intercal_\G A \gamma_\G$  with $\gamma = \beta^{(1)} - \beta^{(0)}$ & Generate a Dist object.  \\
        ci() & & Input object, return CIs. \\
        summary() & & Input object, compute and return a list of summary statistics, including bias-corrected point estimators, standard error and so on. \\
        \bottomrule
    \end{tabular}
    }
    \caption{Functions of \CRANpkg{SIHR}}
    \label{tab: overview_package}
\end{table}

\subsection{Linear functional}
\label{subsec: LF}
The function \code{LF()}, shorthanded for Linear Functional, performs inference for $\xnew^{\intercal}\beta$ under the high-dimensional model \eqref{eq: glm}.  A typical \code{LF()} code snippet looks like:\\
\\
\code{
LF(X, y, loading.mat, model=c("linear","logistic","logistic\_alter"), intercept=TRUE, \\
\indent intercept.loading=FALSE, beta.init=NULL, lambda=NULL, mu=NULL, prob.filter=0.05, \\
\indent rescale=1.1, alpha=0.05, verbose=FALSE)
}\\

 The argument \code{loading.mat} takes values of $\xnew$ as a matrix, which allows for multiple $\xnew$ as the input, with each column representing a new future observation $\xnew \in \RR^p$. The argument \code{model} specifies what regression model the algorithm is working on, which can take ``linear'', ``logistic'', ``logistic\_alter'', corresponding to the Table \ref{tab: components}. The argument \code{intercept.loading} is logical, specifying whether the intercept term should be included or not for defining the objective $\xnew^{\intercal}\beta$ and the default value is FALSE. More detailed descriptions of each input argument in \code{LF()} function can be found in Table \ref{tab: details_LF}. In the following code, we make inference for $\xnew^\intercal \beta$ with simulated data when the outcome $y_i$ is continuous.

\begin{table}[ht]
    \centering
    \small
    \resizebox{0.9\textwidth}{!}{
    \begin{tabular}{m{2.2cm} m{10cm} m{2cm}}
        \toprule
        Argument & Description & Default\\
        \midrule
        X & Design matrix, of dimension $n \times p $ &  \\
        y & Outcome vector, of length $n$ &  \\
        loading.mat & Loading matrix each column corresponds to a loading of interest $\xnew$ & \\
        model & The regression model to fit, one of ``linear'', ``logistic'' and ``logistic\_alter'' & ``linear'' \\
        intercept & Should intercept be fitted for the initial estimator & TRUE\\
        intercept.loading & Should intercept term be included for the inference of objective. & FALSE \\
        beta.init & The initial estimator of the regression vector & NULL\\
        lambda & The tuning parameter in fitting initial model. If not specified, it will be picked by cross-validation. & NULL\\
        mu & The dual tuning parameter used in the construction of the projection direction. If not specified, it will be searched automatically. & NULL\\
        prob.filter & The threshold of estimated probabilities for filtering observations for binary outcome. & 0.05 \\
        rescale & The factor to enlarge the standard error to account for the finite sample bias. & 1.1\\
        alpha & Level of significance to construct two-sided CI & 0.05 \\
        verbose & Should intermediate message(s) be printed, the projection direction be returned. & FALSE\\
        \bottomrule
    \end{tabular}}
    \caption{Arguments for function \code{LF()}}
    \label{tab: details_LF}
\end{table}

\noindent\textbf{Example 1.} For $1\leq i\leq n$, the covariates $X_{i}$ are independently generated from the multivariate normal distribution with mean $\mu = 0_p$ and covariance $\Sigma = \mathbf{I}_p$. The outcome is generated as $y_i = a_0 + X_{i\cdot}^\intercal \beta + \epsilon_i$ with standard normal noise.
Given two further observations $\xnew^{(1)}, \xnew^{(2)}$, we're going to make inference for $\xnew^{(1)\intercal}\beta$ and $\xnew^{(2)\intercal}\beta$ simultaneously.
\begin{example}
## Data Preparation ##
set.seed(0)
n = 100; p = 120
mu = rep(0,p); Cov = diag(p)
a0 = -0.5
beta = rep(0,p); beta[c(1,2)] = c(0.5, 1)
X = MASS::mvrnorm(n, mu, Cov)
y = a0 + X 

## two further observations ##
loading1 = c(1, 1, rep(0, p-2))
loading2 = c(-0.5, -1, rep(0, p-2))
loading.mat = cbind(loading1, loading2)

## Linear Functional ##
Est = LF(X, y, loading.mat, model='linear')
\end{example}
Having fitted the model, we have two following functions \code{ci()} and \code{summary()}.
\begin{example}
ci(Est)
#>  loading     lower      upper
#>1       1  1.167873  1.8753934
#>2       2 -1.544138 -0.7995375
\end{example}
In the above result, we can find the $95\%$ CI for $\xnew^{(1)\intercal}\beta$ and $\xnew^{(2)\intercal}\beta$. Both true values $\xnew^{(1)\intercal}\beta = 1.5$ and $\xnew^{(2)\intercal}\beta = -1.25$ lie in the corresponding CIs.
\begin{example}
summary(Est)
#>Call: 
#>Inference for Linear Functional
#>
#>Estimators: 
#> loading est.plugin est.debias Std. Error z value  Pr(>|z|)   
#>       1      1.268      1.522     0.1805   8.430 0.000e+00 ***
#>       2     -1.033     -1.172     0.1900  -6.169 6.868e-10 ***
\end{example}
\code{summary()} returns a list of the summary statistics, in which we can find the plugin estimator, bias-corrected estimator, and the standard error for the bias-corrected estimator. The bias-corrected estimators are closer to the true values.

As a second example, we consider the logistic regression where the argument \code{model} is set as \code{"logistic"} or \code{"logistic\_alter"}. To boost computation efficiency, we may specify the argument \code{beta.init} as the common initial coefficients estimators for all further observations.

\noindent\textbf{Example 2.} For $1\leq i\leq n$, the covariates $X_{i}$ are independently generated from the multivariate normal distribution with mean $\mu = 0_p$ and covariance $\Sigma = \mathbf{I}_p$. We generate the outcome following the model $Y_{i} \sim {\rm Bernoulli}\left({f\left(a_0 + X_{i}^{\intercal}\beta\right)}\right)$
with $f(z) = \exp(z)/[1+\exp(z)]$.

\begin{example}
## Data Preparation ##
set.seed(0)
n = 300; p = 120
mu = rep(0,p); Cov = diag(p)
a0 = -1
beta = rep(0,p); beta[c(1,2)] = c(1, 1)
X = MASS::mvrnorm(n, mu, Cov)
val = a0 + X 
y = rbinom(n, 1, exp(val)/(1+exp(val)))

## two further observations ##
loading1 = c(1, 1, rep(0, p-2))
loading2 = c(-0.5, -2, rep(0, p-2))
loading.mat = cbind(loading1, loading2)

## obtain initial estimators ##
cv.fit = glmnet::cv.glmnet(X, y, family='binomial', alpha=1, standardize=TRUE)
beta.init = as.vector(coef(cv.fit, s=cv.fit[['lambda.min']]))
Est = LF(X, y, loading.mat, model='logistic', beta.init=beta.init) 
\end{example}
The corresponding CIs and summary statistics are given below:
\begin{example}
ci(Est)
#>  loading      lower      upper
#>1       1  1.257559  2.492327
#>2       2 -3.186513 -1.605671
\end{example}
Consequently, we have two objective values $\xnew^{(1)\intercal}\beta = 2$ and $\xnew^{(2)\intercal}\beta = -2.5$. Both of these values lie within their corresponding 95\% CIs.
\begin{example}
summary(Est)
#> Call: 
#> Inference for Linear Functional
#> 
#> Estimators: 
#> loading est.plugin est.debias Std. Error z value  Pr(>|z|)    
#>       1      1.340      1.875     0.3150   5.952 2.645e-09 ***
#>       2     -1.741     -2.396     0.4033  -5.941 2.825e-09 ***
\end{example}
Note that the plugin estimators $\xnew^{(1)\intercal}\widehat{\beta}$ and $\xnew^{(2)\intercal}\widehat{\beta}$ are severely biased in such setting, the proposed bias-correction approach significantly saves the bias with $\widehat{\xnew^{(1)\intercal}{\beta}}$ and $\widehat{\xnew^{(2)\intercal}{\beta}}$. 

\subsection{Quadratic functional}
\label{subsec: QF}
The function \code{QF()}, abbreviated for Quadratic Functional, conducts inference for $\beta_{\textrm{G}}^{\intercal}A\beta_{\textrm{G}}$ if $A$ is the submatrix pre-specified or $\beta_{\textrm{G}}^\intercal \Sigma_{\textrm{G,G}} \beta_{\textrm{G}} $ under the high-dimensional regression model \eqref{eq: glm}. The function \code{QF()} can be called with the following arguments.\\
\\
\code{QF(X, y, G, A=NULL, model=c("linear","logistic","logistic\_alter"), intercept=TRUE,\\
\indent beta.init=NULL, split=TRUE, lambda=NULL, mu=NULL, prob.filter=0.05, rescale=1.1,\\
\indent tau=c(0.25, 0.5, 1), alpha=0.05, verbose=FALSE)}
\\

The argument \code{G} is the set of indices of interest. If the argument \code{A} is specified, it will conduct inference for $\beta_{\textrm{G}}^\intercal A \beta_{\textrm{G}}$; otherwise, it will turn to $\beta_{\textrm{G}}^\intercal \Sigma_{{\textrm{G,G}}} \beta_{\textrm{G}}$. The argument \code{model} specifies what regression model the algorithm is working on, which can take ``linear'', ``logistic'', ``logistic\_alter'', corresponding to Table \ref{tab: components}. The argument \code{split} indicates whether we conduct the sample splitting for computing the initial estimator of regression coefficients. {When \code{split=TRUE}, the initial estimator of regression coefficients is computed using half of the available observations while the remaining half is used for bias correction. The option of using sampling splitting might require a larger sample size.} The argument \code{tau.vec} allows the user to supply a vector of possible values for $\tau$ in (\ref{eq: QF CI1 - unifed}) and \eqref{eq: QF CI2 - unified}. 
Table \ref{tab: details_QF} presents the arguments for \code{QF()} while excluding those that are repeated in Table \ref{tab: details_LF}. 

\begin{table}[ht]
    \centering
    \small
    \resizebox{0.9\textwidth}{!}{
    \begin{tabular}{m{2.2cm} m{10cm} m{2cm}}
        \toprule
        Argument & Description & Default\\
        \midrule
        X & Design matrix, of dimension $n \times p $ &  \\
        y & Outcome vector, of length $n$ &  \\
        G & The set of indices in the quadratic form & \\
        A & The matrix A in the quadratic form, of dimension $|G| \times |G|$. If not specified, A would be set as the $|G|\times |G|$ submatrix of the population covariance matrix corresponding to the index set G & NULL\\
        model & The regression model to fit, one of ``linear'', ``logistic'' and ``logistic\_alter'' & ``linear''\\
        intercept & Should intercept be fitted for the initial estimator & TRUE\\
        split & Sampling splitting or not for computing the initial estimator. It takes effect only when beta.init =  NULL. & TRUE \\
        tau & The enlargement factor for asymptotic variance of the bias-corrected estimator to handle super-efficiency. It allows for a scalar or vector. & c(0.25, 0.5, 1) \\
        \bottomrule
    \end{tabular}
    }
    \caption{Arguments for function \code{QF()}, others are repeated in Table \ref{tab: details_LF}}
    \label{tab: details_QF}
\end{table}

In the third example, we illustrate the usage of \code{QF()} in linear regression model,

\noindent\textbf{Example 3.} For $1\leq i\leq n$, the covariates $X_{i\cdot}$ is generated from multivariate normal distribution with mean $\mu = 0_p$ and covariance $\Sigma \in \mathbb{R}^{p \times p}$ where $\Sigma_{j, k} = 0.5^{|j-k|}$. 
We generate the outcome following the model $y_i = X_{i\cdot} \beta + \epsilon_i$ with standard normal distributed noise. We're going to make inference for $\beta_{\G}^{\intercal} \Sigma_{\G,\G} \beta_\G$ with $G=\{40,\ldots,60\}$.

\begin{example}
## Data Preparation ##
set.seed(0)
n = 200; p = 150
mu = rep(0,p)
Cov = matrix(0, p, p); for(j in 1:p) for(k in 1:p) Cov[j,k] = 0.5^{abs(j-k)}
beta = rep(0, p); beta[25:50] = 0.2
X = MASS::mvrnorm(n,mu,Cov)
y = X

## set G ##
test.set =c(40:60)

## Quadratic Functional ##
Est = QF(X, y, G = test.set, A = NULL, model = "linear", split=FALSE)
\end{example}

Continuing running two functions \code{ci()} and \code{summary()}:
\begin{example}
ci(Est)
#>   tau     lower    upper
#>1 0.25 0.8118792 1.466422
#>2 0.50 0.8046235 1.473677
#>3 1.00 0.7905648 1.487736
\end{example}
With the default $\tau=c(0.25, 0.5, 1)$, we obtain three different CIs for $\beta_{\G}^{\intercal} \Sigma_{\G,\G} \beta_\G$. Note that the true value $\beta_{\G}^{\intercal}\Sigma_{\G,\G}\beta_{\G} = 1.16$ belongs to all of these constructed CIs.
\begin{example}
summary(Est)
#> Call: 
#> Inference for Quadratic Functional
#> 
#>  tau est.plugin est.debias Std. Error z value  Pr(>|z|)    
#> 0.25      0.904      1.139     0.1670   6.822 8.969e-12 ***
#> 0.50      0.904      1.139     0.1707   6.674 2.486e-11 ***
#> 1.00      0.904      1.139     0.1779   6.405 1.504e-10 ***
\end{example}
Similarly to the \code{LF()} case, our proposed bias-corrected estimator is effective in correcting the bias of plugin estimator.

\subsection{Conditional average treatment effect}
\label{subsec: CATE}
The function \code{CATE()}, shorthanded for Conditional Average Treatment Effect, conducts inference for $\Delta(\xnew) = f(\xnew^{\intercal}\beta^{(2)})-f(\xnew^\intercal \beta^{(1)})$ under the high-dimensional regression model \eqref{eq: two sample glm}. 
This function can be implemented as follows:

\code{CATE(X1, y1, X2, y2, loading.mat, model=c("linear","logistic","logistic\_alter"),\\
\indent intercept=TRUE, intercept.loading=FALSE, beta.init1=NULL, beta.init2=NULL, \\
\indent lambda=NULL, mu=NULL, prob.filter=0.05, rescale=1.1, alpha=0.05, verbose=FALSE)}
\\

Here, \code{X1} and \code{y1} denote the design matrix and the response vector for the first sample of data respectively, while \code{X2} and \code{y2} denote those for the second sample of data. \code{beta.init1} and \code{beta.init2} are the initial estimator of the regression vector for the first and second samples. All other arguments are similarly defined as for the function \code{LF()}.

For the fourth example, we consider the logistic regression case to illustrate \code{CATE()} with the argument 
\code{model='logistic\_alter'}.

\noindent\textbf{Example 4.}
In the first group of data, the covariates $X_{i\cdot}^{(1)}$ follows multivariate normal distribution with $\mu = 0_p$ and covariance $\Sigma=\mathbf{I}_p$; in the second group of data, the covariates $X_{i\cdot}^{(2)}$ follows multivariate normal distribution with $\mu=0_p$ and covariance $\Sigma \in \RR^{p\times p}$ with $\Sigma_{j,k} = 0.5^{|j-k|}$. We generate following the model $y_i^{(k)} \sim \textrm{Bernoulli}(f(X_{i\cdot}^{(k)\intercal}\beta^{(k)})$ with $f(z) = \exp(z)/[1+\exp(z)]$ for $k=1,2$. See the following code for details of $\beta^{(1)}$, $\beta^{(2)}$ and the further observation $\xnew$.

    
\begin{example}
## Data Preparation ##
set.seed(0)
n1 = 100; n2 = 180; p = 120
mu1 = mu2 = rep(0,p)
Cov1 = diag(p)
Cov2 = matrix(0, p, p); for(j in 1:p) for(k in 1:p) Cov2[j,k] = 0.5^{abs(j-k)}
beta1 = rep(0, p); beta1[c(1,2)] = c(0.5, 0.5)
beta2 = rep(0, p); beta2[c(1,2)] = c(1.8, 1.8)
X1 = MASS::mvrnorm(n1,mu1,Cov1); val1 = X1
X2 = MASS::mvrnorm(n2,mu2,Cov2); val2 = X2
y1 = rbinom(n1, 1, exp(val1)/(1+exp(val1)))
y2 = rbinom(n2, 1, exp(val2)/(1+exp(val2)))

## further observation ##
loading.mat = c(1, 1, rep(0, p-2))

## CATE ##
Est <- CATE(X1, y1, X2, y2,loading.mat, model="logistic_alter")
\end{example}

\noindent Having fitted the model, it allows for method \code{ci()} and \code{summary()} as \code{LF()} does.
\begin{example}
ci(Est)
#>  loading    lower    upper
#>1       1 1.614269 4.514703
\end{example}
The true value $\xnew^\intercal (\beta^{(2)} - \beta^{(1)}) = 2.6$ is included in the above $95\%$ CI.
\begin{example}
ci(Est, probability = TRUE)
#>  loading     lower     upper
#>1       1 0.1531872 0.5086421
\end{example}
If we specify \code{probability} as \code{TRUE}, for the logistic regression, \code{ci()} yields the CI for $f(\xnew^\intercal \beta^{(2)}) - f(\xnew^\intercal \beta^{(1)})$ whose true value is $0.2423$.

\subsection{Inner Product}
\label{subsec: InnProd}
The function \code{InnProd()}, shorthanded for Inner Product, conducts inference for $\beta_{\textrm{G}}^{(1)\intercal}A\beta_{\textrm{G}}^{(2)}$ if $A$ is the submatrix pre-specified or $\beta_{\textrm{G}}^{(1)\intercal} \Sigma_{\textrm{G,G}} \beta_{\textrm{G}}^{(2)} $ under the high-dimensional regression models. Here, \code{X1} and \code{y1} denote the design matrix and the response vector for the first sample of data respectively, while \code{X2} and \code{y2} denote those for the second sample of data. All other arguments are similarly defined as for the function \code{QF()}. 
\\
\\
\code{
InnProd(X1, y1, X2, y2, G, A = NULL, model=c("linear","logistic","logistic\_alter"), intercept=TRUE, \\
\indent beta.init1=NULL, beta.init2=NULL, split = TRUE,lambda=NULL, mu=NULL, prob.filter=0.05, \\
\indent rescale=1.1, tau = c(0.25,0.5,1), alpha=0.05, verbose=FALSE)
}\\

\noindent In the following code, we demonstrate the use of \code{InnProd()} in linear regression.\\
\noindent\textbf{Example 5.} See the following code for generating two samples of data and inference for $\beta_{\textrm{G}}^{(1)\intercal}A\beta_{\textrm{G}}^{(2)}$.

\begin{example}
set.seed(0)
n1 = 200; n2 = 260; p = 120
mu1 = mu2 = rep(0,p)
Cov1 = diag(p)
Cov2 = matrix(0, p, p); for(j in 1:p) for(k in 1:p) Cov2[j,k] = 0.5^{abs(j-k)}
beta1 = rep(0, p); beta1[1:10] = 0.5
beta2 = rep(0, p); beta2[3:12] = 0.4
X1 <- MASS::mvrnorm(n1,mu1,Cov1)
X2 <- MASS::mvrnorm(n2,mu2,Cov2)
y1 <- X1
y2 <- X2

## Specify G and A ##
test.set = c(1:20)
A = diag(length(test.set))

## Inner Product ##
Est <- InnProd(X1, y1, X2, y2, G=test.set, A, model="linear")
\end{example}

Having fitted the model, it allows for method \code{ci()} and \code{summary()} as \code{QF()} does.
\begin{example}
ci(Est)
#>    tau     lower    upper
#> 1 0.25 0.7432061 2.490451
#> 2 0.50 0.7128181 2.520839
#> 3 1.00 0.6520422 2.581615
\end{example}
The true value $\beta^{(1)\intercal} A \beta^{(2)} = 1.6$ is included in the above CIs with all default $\tau$ values. 

\subsection{Distance}
\label{subsec: Dist}
The function \code{Dist()}, shorthanded for Distance, conducts inference for $\gamma_{\textrm{G}}^{\intercal}A\gamma_{\textrm{G}}$, where $\gamma = \beta^{(1)}-\beta^{(0)}$, if $A$ is the submatrix pre-specified or $\gamma_{\textrm{G}}^{\intercal}\Sigma_{\mathrm{G},\mathrm{G}}\gamma_{\textrm{G}}$ under the high-dimensional regression models. All arguments are similarly defined as for the function \code{InnProd()}. 
\\
\\
\code{
Dist(X1, y1, X2, y2, G, A = NULL, model=c("linear","logistic","logistic\_alter"), intercept=TRUE, \\
\indent beta.init1=NULL, beta.init2=NULL, split = TRUE, lambda=NULL, mu=NULL, prob.filter=0.05, \\
\indent rescale=1.1, tau = c(0.25,0.50,1), alpha=0.05, verbose=FALSE)
}\\

\noindent In Example 6 we illustrate the use of \code{Dist()} in linear regression.\\
\noindent\textbf{Example 6.} 
See the following code for generating two samples of data and inference for $\gamma_{\textrm{G}}^{(1)\intercal}\Sigma_{\G,\G}\gamma_{\textrm{G}}^{(2)}$.

\begin{example}
## Data Preparation ##
set.seed(0)
n1 = 220; n2 = 180; p = 100
mu = rep(0,p); Cov = diag(p)
beta1 = rep(0, p); beta1[1:2] = c(0.5, 1)
beta2 = rep(0, p); beta2[1:10] = c(0.3, 1.5, rep(0.08, 8))
X1 <- MASS::mvrnorm(n1,mu,Cov)
X2 <- MASS::mvrnorm(n2,mu,Cov)
y1 = X1
y2 = X2

## G ##
test.set = c(1:10)

## A is not specified $$$$
Est <- Dist(X1, y1, X2, y2, G=test.set, A=NULL, model="linear", split=FALSE)
\end{example}

Having fitted the model, it allows for method \code{ci()} and \code{summary()} as \code{LF()} does.
\begin{example}
ci(Est)
#>   tau    lower     upper
#>1 0.25 0.028202 0.6831165
#>2 0.50 0.000000 0.7196383
#>3 1.00 0.000000 0.7926819

summary(Est)
#> Call: 
#> Inference for Distance
#> 
#>  tau est.plugin est.debias Std. Error z value Pr(>|z|)  
#> 0.25     0.4265     0.3557     0.1671   2.129  0.03327 *
#> 0.50     0.4265     0.3557     0.1857   1.915  0.05547 .
#> 1.00     0.4265     0.3557     0.2230   1.595  0.11070  
\end{example}
The true value $\gamma_\G^\intercal \Sigma_{\G, \G} \gamma_\G = 0.3412$. Similar to the previous instances, we note that the bias-corrected estimator effectively correct the bias of the plugin estimator. Depending on the $\tau$ values, we obtain various CIs, all of which encompass the true value. It is important to mention that in case of negative lower boundaries, they will be truncated at $0$ for $\tau=0.5$ and $\tau=1$.

\section{Applications}
\label{sec: real data}
\subsection{Motif Regression}
We demonstrate the use of \code{LF()} function on a motif regression problem for predicting transcription factor binding sites (TFBS, also called `motifs') in DNA sequences. The data set consists of a univariate response variable $y$ measuring the binding intensity of the transcription factor on coarse DNA segments for $n = 2587$ genes. Moreover, for each of the $n$ genes, a score describing the abundance of occurrence, is available for each of the $p = 666$ candidate motifs. This data set has been previously explored in \citet{motif}. To summarize, we have the following data:
$$
\begin{aligned}
y_i : & \text{the binding intensity of the transcription factor on coarse DNA segment } i\\ 
X_{i,j} : & \text{the abundance score of candidate motif } j \text{ in DNA segment } i \\
i = & 1,\cdots,n;\quad j = 1,\cdots,p
\end{aligned}
$$
Given the real data, we run the following code:
\begin{example}
    p = ncol(X)
    loading.mat = diag(p)
    ## apply LF ##
    Est = LF(X, y, loading.mat, model='linear')
    ## CI for each regression coef ##
    ci(Est)
\end{example}



\begin{figure}[htp]
\centering
\includegraphics[width = \textwidth]{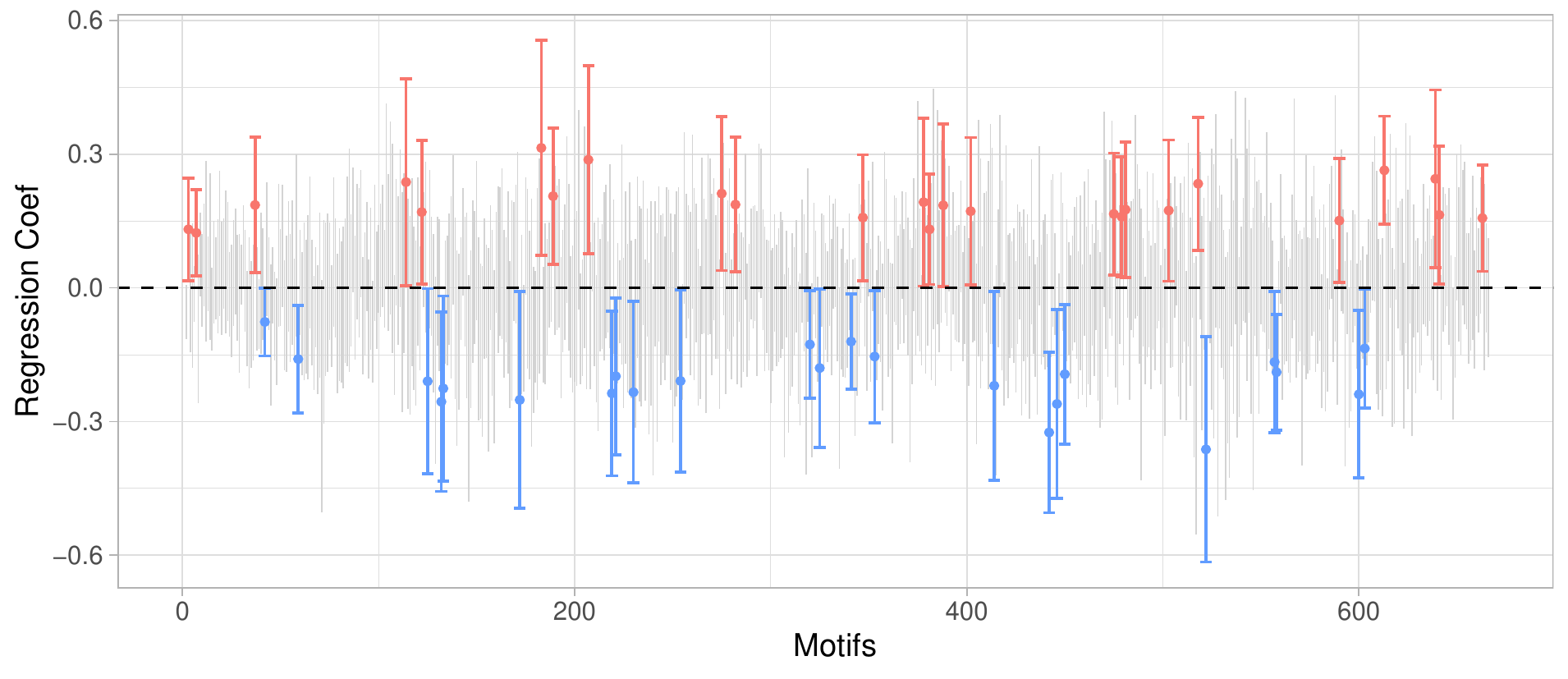}
\caption{Constructed CIs for the 666 regression coefficients.} 
\label{fig: t134}
\end{figure}

We apply the package function \code{LF()} and obtain $95 \%$ CIs for the 666 regression coefficients. The constructed CIs are illustrated in Figure \ref{fig: t134}. 
{Among the 666 CIs, 25 are marked in red and lie completely above $0$, suggesting a positive relationship between the Motif and the binding intensity. Conversely, 23 of the CIs highlighted in blue lie completely below $0$, indicating that the corresponding motif has a negative impact on the binding intensity of the transcription factor.}
In other words, many genes may be targeted by the transcription factors that bind to these 48 motifs.


\subsection{Fasting Glucose Level Data}
The aim is to analyze the effect of polymorphic genetic markers on the glucose level in a stock mice population using the \code{LF()} with the argument \code{model="logistic"}. The data set is available at \url{https://wp.cs.ucl.ac.uk/outbredmice/heterogeneous-stock-mice/.} Since fasting glucose level is an important indicator of type$-2$ diabetes, the fasting glucose level dichotomized at $11.1$ (unit: mmol/L) is taken as the response variable. Specifically, glucose level below $11.1$ is considered normal and above $11.1$ high (pre-diabetic and diabetic). The covariates consist of $10,346$ polymorphic genetic markers, and the sample size is $1,269$. We include ``gender" and ``age" as baseline covariates. The polymorphic markers and baseline covariates are standardized before analysis. However, the number of polymorphic markers is large, and there exists a high correlation among some of them. To address this issue, we select a subset of polymorphic markers such that the maximum of absolute correlation among the markers is below $0.75$. 
Eventually, we select a subset of $2, 341$ polymorphic markers. 
{To sum up, we have the following data,  for $i = 1,\cdots,1269$:
$$
\begin{aligned}
y_i : & \text{ whether the fasting glucose level is above $11.1$ mmol/L for unit} i \\
X_{i,j} : & \text{ polymorphic marker $j$ for unit } i\; \textrm{ with $j=1,2,...,2341$} \\
X_{i, 2342} : & \text {gender of unit } i \\
X_{i,2343} : & \text{ age of unit } i
\end{aligned}
$$
Given the real data, we run the following code:}
\begin{example}
    p = ncol(X)
    loading.mat = diag(p)[,-c(2342,2343)]
    ## apply LF ##
    Est = LF(X, y, loading.mat, model='logistic')
    ## CI for each regression coef ##
    ci(Est)
\end{example}

\begin{figure}[htp]
\centering
\includegraphics[width = \textwidth]{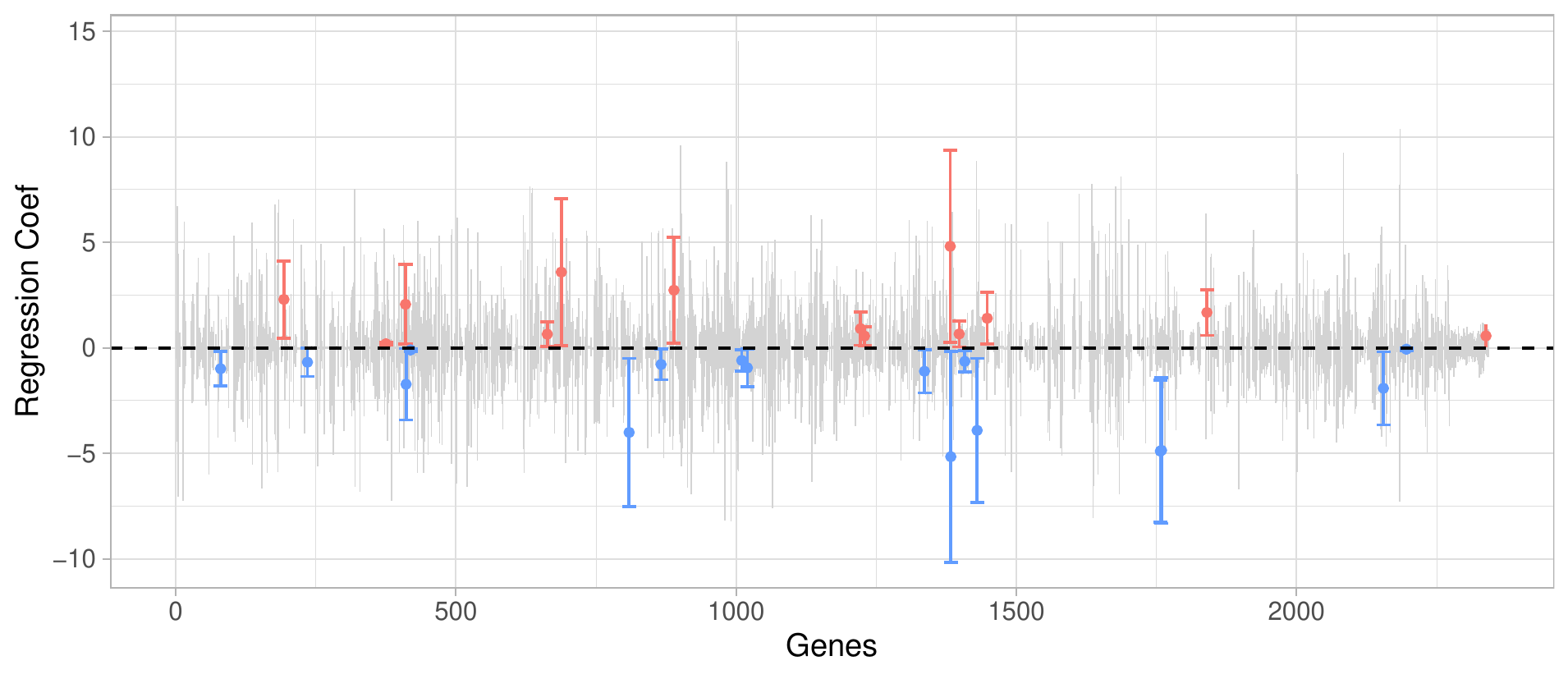}
\caption{Constructed CIs for the $2341$ regression coefficients. 
} 
\label{fig: logistic CI}
\end{figure}

{
Once more, we utilize the package function \code{LF()} with \code{model = "logistic"} to generate CIs for the first 2341 regression coefficients (corresponding to all polymorphic markers). In Figure \ref{fig: logistic CI}, we observe that 13 genes have CIs that lie entirely above $0$ (highlighted in red), while 16 genes have CIs below $0$ (highlighted in blue). This indicates their respective associations with the fasting glucose level.}

\section{Conclusion}


{There has been significant recent progress in debiasing inference methods for high-dimensional GLMs. This paper highlights the application of advanced debiasing techniques in high-dimensional GLMs using the R package \CRANpkg{SIHR}. The package provides tools for estimating bias-corrected point estimators and constructing CIs for various low-dimensional objectives in both one- and two-sample regression settings. Through extensive simulations and real-data analyses, we demonstrate the practicality and versatility of the package across diverse fields of study, making it an essential addition to the literature.
}
\section{Acknowledgement}
Prabrisha Rakshit and Zhenyu Wang contributed equally to this work and are considered co-first authors. Dr. Tony Cai’s research was supported in part by NSF grant DMS-2015259 and NIH grants R01-GM129781 and R01-GM123056. Dr. Zijian Guo’s research was supported in part by NSF grants DMS-1811857 and DMS-2015373 and NIH grants R01-GM140463 and R01-LM013614. Dr. Zijian Guo is grateful to Dr. Lukas Meier for sharing the motif regression data used in this paper.

\bibliography{rakshit-wang}

\address{Prabrisha Rakshit\\
  Rutgers, The State University of New Jersey\\
  Address\\
  USA\\
  \email{prabrisha.rakshit@rutgers.edu}}

\address{Zhenyu Wang\\
  Rutgers, The State University of New Jersey\\
  Address\\
  USA\\
  \email{zw425@stat.rutgers.edu}}

\address{Tony Cai\\
  University of Pennsylvania\\
  Address\\
  USA\\
  \email{tcai@wharton.upenn.edu}}
  
\address{Zijian Guo\\
  Rutgers, The State University of New Jersey\\
  Address\\
  USA\\
  \email{zijguo@stat.rutgers.edu}}  

\end{article}

\end{document}